\documentclass[3p,english,12pt,times,sort&compress,nopreprintline]{elsarticle}
\usepackage{fullpage}

\usepackage[utf8]{inputenc}
\usepackage[T1]{fontenc}
\usepackage{babel}
\usepackage{array}
\usepackage{multirow}
\usepackage{amsmath}
\usepackage{amsthm}
\usepackage{amssymb}
\usepackage{nicefrac}
\usepackage{graphicx}
\usepackage[unicode=true]{hyperref}
\usepackage{xcolor}
\usepackage{textgreek}
\usepackage{comment}
\usepackage[title]{appendix}

\hypersetup{
  colorlinks=true,
  linkcolor=blue,
  filecolor=magenta,
  urlcolor=cyan,
}

\makeatletter
\newcommand{\orcidID}[1]{\href{https://orcid.org/#1}{\includegraphics[width=10pt]{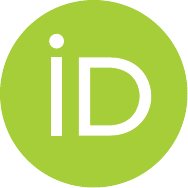}}}
\providecommand{\tightlist}{\setlength{\itemsep}{0pt}\setlength{\parskip}{0pt}}
\def\autocite#1\citep{#1}
\theoremstyle{plain}

\theoremstyle{definition}

\theoremstyle{remark}

\theoremstyle{plain}

\makeatother

\providecommand{\definitionname}{Definition}
\providecommand{\lemmaname}{Lemma}
\providecommand{\remarkname}{Remark}
\providecommand{\theoremname}{Theorem}

\makeatletter
\def\maxwidth{\ifdim\Gin@nat@width>\linewidth\linewidth\else\Gin@nat@width\fi}
\def\maxheight{\ifdim\Gin@nat@height>\textheight\textheight\else\Gin@nat@height\fi}
\makeatother
\setkeys{Gin}{width=\maxwidth,height=\maxheight,keepaspectratio}

\begin{document}

\begin{frontmatter}

\title{Dyadic aggregated autoregressive (DASAR) model for time-frequency
representation of biomedical signals}

\author[mek]{Marco A. Pinto Orellana \orcidID{0000-0001-6495-1305}}
\corref{cor}
\author[mek]{Habib Sherkat }
\author[mek]{Peyman Mirtaheri \orcidID{0000-0002-7664-5513}}
\author[it,simulamet]{Hugo L. Hammer \orcidID{0000-0001-9429-7148}}

\cortext[cor]{Corresponding author}

\address[mek]{Department of Mechanical, Electronics and Chemical
Engineering. Oslo Metropolitan University.}
\address[it]{Department of Information Technology. Oslo Metropolitan
University.}
\address[simulamet]{Department of Holistic Systems, Simula Metropolitan
Center for Digital Engineering.}

\begin{abstract}
This paper introduces a new time-frequency representation method for
biomedical signals: the dyadic aggregated autoregressive (DASAR) model.
Signals, such as electroencephalograms (EEGs) and functional
near-infrared spectroscopy (fNIRS), exhibit physiological information
through time-evolving spectrum components at specific frequency
intervals: 0-50 Hz (EEG) or 0-150 mHz (fNIRS). Spectrotemporal features in
signals are conventionally estimated using short-time Fourier transform
(STFT) and wavelet transform (WT).
However, both methods may not offer the most robust or compact representation
despite their widespread use in biomedical contexts. 
The presented method, DASAR, improves precise frequency
identification and tracking of interpretable frequency components with a
parsimonious set of parameters. DASAR achieves these characteristics by
assuming that the biomedical time-varying spectrum comprises
several independent stochastic oscillators with (piecewise) time-varying
frequencies. Local stationarity can be assumed within dyadic
subdivisions of the recordings, while the stochastic oscillators can be
modeled with an aggregation of second-order autoregressive models (ASAR).
DASAR can provide a more accurate representation of the 
(highly contrasted) EEG and fNIRS frequency ranges by increasing the estimation
accuracy in user-defined spectrum region of interest (SROI).
A mental arithmetic experiment on a hybrid EEG-fNIRS was conducted to assess 
the efficiency of the method.
Our proposed technique, STFT, and WT were applied on both biomedical signals
to discover potential oscillators that improve the discrimination between the task 
condition and its baseline. The results show that DASAR provided the highest spectrum 
differentiation and it was the only method
that could identify Mayer waves as narrow-band artifacts at
97.4-97.5 mHz.
\end{abstract}


\end{frontmatter}

\section{Introduction}

The brain is a complex and strongly interconnected organ with several electrical and biochemical interactions between its cellular constituents
\citep{DensityNeuronsSynapses-Schuz-1989, MorphologicalDistribution-Reina-DeTorre-1998, NeuronalBloodbrain-Kaplan-2020}.
These interactions involve biological structures and functionalities,
including vessels, glial cells (astrocytes, oligodendrocytes, and microglial), and neurons, explained by the
neurovascular coupling theory \citep{NeuronalBloodbrain-Kaplan-2020, NewCognitiveNeuroscience-Gramann-2014}.
As a result, these brain interactions produce biomedical signals
with time- and frequency-varying properties, which can be applied for monitoring brain's pathophysiology \citep{BrainChirpsSpectrographic-Schiff-2000} or cognitive
loading \citep{BrainOscBiomarkers-Yener-2013}.

This paper, focuses on the non-invasively measured biomedical time series of
 hemodynamic variations and electrical potentials. In the first category, functional near-infrared
spectroscopy (fNIRS) is used to estimate the concentration of oxygenated
hemoglobin (HbO), and its deoxygenated state (HbR)
\citep{NoninvasiveOptical-Villringer-1997}.
The fNIRS imaging technique  uses absorbance of near-infrared (NIR) light in the blood to estimate the hemoglobin
concentration. The notable  difference of HbR and HbO wavelength
absorption properties in the NIR region (750-850 nm) allows us to obtain a proper
concentration estimation of both chromophores based on the
modified Beer-Lambert law \citep{NoninvasiveOptical-Villringer-1997, HandbookNIR-Burns-2008, NoninvasiveInfraredMonitoring-Jobsis-1977}.
Electroencephalograms (EEGs),  the second modality, record the
electrical potential difference at several cranial locations
\citep{AtlasEEGPatterns-Stern-2013}. Positive and negative charges stem as propagation of electrical waves produced by large synchronized groups of pyramidal neurons in the brain's local cortical regions \citep{AtlasEEGSeizure-Khalil-2006}.

Combined hemodynamic-electrical modality enhances brain-computer interface as it reflects neuron firing activities whiles provide information about oxygen/energy consumption in the brain's region of interest \citep{NewCognitiveNeuroscience-Gramann-2014}. Thus it will uncover high
nonlinear relationships between recordings and stimuli
\citep{HybridBrainComputerInterface-Hong-2017}, which may result in better understanding of the working mechanisms of the brain under some diseases
\citep{DetectionHemoResponses-Machado-2011}. \\In both 
signals, features or properties extracted from time-frequency
characteristics are usually conditioned on external stimuli and are used
to differentiate activities. However, the time-frequency attributes
are biologically distinct for the EEG and fNIRS. As an example, a common
frequency fNIRS bandwidth is normally described in the range of very
low-frequency waves: 10-100 mHz \citep{CurrentStatusIssues-Pinti-2019}. A
typical hemodynamic response has a frequency of approximately 45 mHz
($\nicefrac{1}{22\text{s}}$)
\citep{StatisticalParamMapping-Penny-2011} or 78 mHz
($\nicefrac{1}{12.75\text{s}}$) during sensorimotor activities
\citep{DeconvolutionImpulseResponse-Glover-1999, AberrantHemoResponses-Yan-2018}.
On the other hand, electroencephalogram waves have a range on a broader
frequency intervals: 0-50 Hz. The EEG complete spectrum is
divided into five major ranges: delta (0-4 Hz), theta (4-8 Hz),
alpha (8-12 Hz), beta (12-30 Hz), and gamma rhythms (\textgreater{} 30 Hz)
\citep{AtlasEEGPatterns-Stern-2013, ElecFieldsBrain-Nunez-2006}.

Consequently, two natural frameworks that would address the extraction
of time-varying spectral information of biomedical
signals are the short-time Fourier transform (STFT) and wavelet transform (WT).
STFT has been reportedfor extracting patterns to measure mental workload
\citep{MentalWorkload-Aghajani-2017}, epilepsy
\citep{EpilepticSeizureClassif-Samiee-2015} or autism spectrum disorder
\citep{AnalysesEEGBackground-Behnam-2007} identification, during writing
\citep{STFT-Zabidi-2012, STFT-Fadzal-2012}. While WT has been
employed to obtain features for glucose measurement
\citep{StudyRvwWavelet-Li-2015}, pain assessment
\citep{PainDetectionFNIRS-Lopez-Martinez-2019, CorticalNetworkResponse-Rojas-2019},
artifact correction
\citep{KurtosisbasedWaveletAlgorithm-Chiarelli-2015, AutomaticRemovalEye-Joyce-2004, RemovalOcularArtifact-Kenemans-1991},
epilepsy classification \citep{EEGRecords-Adeli-2003}, or cerebral
autoregulation processes \citep{RvwWaveletTransform-Addison-2015}.

As an alternative time-frequency representation, the dyadic aggregated
autoregressive (DASAR) model, a novel time-frequency approach, was
initially proposed in \citep{DyadicAggregatedAR-Pinto-Orellana-2020} as
a tool to extract evolving frequency properties from modulated signals
as features for automatic classification of modulation. DASAR proposes
an approximation of an evolving spectrum through dyadically splitting a
signal and provides an approximation of the spectrum of each segment
using an aggregated second-order autoregressive model (SAR).

This paper aims to propose the use of the DASAR model as a robust
alternative for the time-spectral characterization of biological
signals, with a special focus on hemodynamic (fNIRS) and electrical
(EEG) signals. We also extend DASAR by defining it within a formal
framework based on local stationary processes, while a
frequency-selectivity feature is introduced, and accompanied by an efficient
estimation algorithm. By virtue of these properties, DASAR can compress the stochastic time-frequency characteristics of a signal while retaining only the strongest and main components (oscillators). In addition, DASAR offers an accurate description of the time-varying
response of biomedical signals compared to WT and STFT, while it is maintaining
its interpretable representation due to the reduced number of model's
parameters.

\subsection{Time-varying spectrum}

A real-valued signal $x\left(t\right)$, which origins rely on a
physiological process, can be assumed to have its statistical
parameters smoothly changing over time under normal conditions.
Therefore, it is reasonable to assume a local stationarity property on
EEG \citep[pp.~50-51]{EEGSigProcessing-Sanei-2007} and fNIRS modeling
\citep{NIRSPackage-Fekete-2011}. Without loss of generality, we can
assume $x\left(t\right)$ (sampled in an time interval
$\left[T_{0},T_{1}\right]$) to be a realization from a zero-mean
stochastic process $X\left(t\right)$ with a conjugate covariance
function
\citep{GeneralizationsCyclostationarySig-Napolitano-2012, AdaptiveCoVarEst-Mallat-1998}%
\begin{equation}%
{
R\left(t_{1},t_{2}\right)=\mathbb{E}\left[X\left(t_{1}\right)\overline{X\left(t_{2}\right)}\right]
}\label{eq:cov}%
\end{equation}%
where $\mathbb{E}\left[\cdot\right]$ is the expectation operator and
$\overline{X\left(\cdot\right)}$ is the conjugate of
$X\left(\cdot\right)$.

We can reformulate $R\left(\cdot\right)$ using the time distance
$\Delta t=t_{2}-t_{1}$ and the midpoint
$\tau=\frac{t_{2}+t_{1}}{2}$:%
\begin{equation}%
{
R\left(t_{1},t_{2}\right)=C\left(\tau,\Delta t \right)
}%
\end{equation}%
Note that $C\left(\Delta t,\tau\right)$ is also called as the
ambiguity function.

Consequently, the process $X\left(t\right)$ is defined as a wide-sense
local stationary process if the covariance function can be expressed as
the product of a normalized stationary covariance function
$r\left(\Delta t\right)$ and non-negative instantaneous signal power
function $q\left(\tau\right)$
\citep{InferenceTimevaryingSig-Anderson-2019}:%
\begin{equation}%
{
C\left(\tau,\Delta t\right)=q\left(\tau\right)r\left(\Delta t\right)
}%
\end{equation}%
Recall that
$q\left(\tau\right)=C\left(\Delta t=0,\tau\right)\,\forall\tau$ and
$r\left(\Delta t=0\right)=1.$

Martin and Flandrin defined the time-varying spectrum as the Fourier
transform of $C\left(\tau,\Delta t\right)$ with respect to
$\upsilon=\Delta t$
\citep{WignerVilleSpectral-Martin-1985}:%
\begin{align}%
S\left(\tau,\omega\right) & =\mathfrak{F}_{\nu\rightarrow\omega}\left\{ C\left(\tau,\upsilon\right)\right\} \left(\tau,\omega\right)=\int_{-\infty}^{\infty}C\left(\tau,\upsilon\right)e^{-j\omega\upsilon}d\upsilon\nonumber \\
& =\int_{-\infty}^{\infty}\mathbb{E}\left[X\left(\tau+\frac{1}{2}\upsilon\right)\overline{X\left(\tau-\frac{1}{2}\upsilon\right)}\right]e^{-j\omega\upsilon}d\upsilon%
\label{eq:tv-spectrum}%
\end{align}%

Note, that the time-varying spectrum can be rearranged as the expected
of the Wigner-Ville distribution $W\left\{ \cdot\right\}$ over the
process $X\left(t\right)$
\citep{WignerVilleSpectral-Martin-1985}:%
\begin{align}%
S\left(\tau,\omega\right) & =\mathbb{E}\left[\int_{-\infty}^{\infty}X\left(\tau+\frac{1}{2}\upsilon\right)\overline{X\left(\tau-\frac{1}{2}\upsilon\right)}e^{-j\omega\upsilon}d\upsilon\right]\nonumber \\
& =\mathbb{E}\left[W\left\{ X\left(t\right)\right\} \right]
\end{align}%

\subsection{Short-time Fourier transform spectrogram}

The short-time Fourier transform (STFT) of a real-valued deterministic
function $h\left(\cdot\right)$, with respect to a window function
$g\left(\cdot\right)$, is defined as
\citep{FoundationsTimeFreq-Grochenig-2001}:%
\begin{equation}%
{
V_{g}\left(\tau,\omega\right)=\int_{-\infty}^{\infty}h\left(t\right)\overline{g\left(t-\tau\right)}e^{-2j\pi t\omega}\,dt
}%
\end{equation}%
and the STFT spectrogram, for deterministic signals, as the squared modulus
of the STFT \citep{FoundationsTimeFreq-Grochenig-2001}:%
\begin{equation}%
{
S\left(\tau,\omega\right)\equiv\left|V_{g}\left(\tau,\omega\right)\right|^{2}
}%
\end{equation}%

Recall that in (stochastic) stationary signals, the spectrum is defined
as the Fourier transform of the autocorrelation function. In a
non-stationary setting, the STFT spectrogram can be understood as the
inverse Fourier transform with respect to $t_{2}$ of the Fourier
transform with respect to $t_{1}$ of the conjugate covariance function
(Equation~\ref{eq:cov}) with respect to $t_{1}$, $t_{2}$
:%
\begin{equation}%
{
S\left(\tau,\omega\right)=\mathfrak{F}_{t_{2}\rightarrow\omega}^{-1}\left\{ \mathfrak{F}_{t_{1}\rightarrow\omega}\left\{ R\left(t_{1},t_{2}\right)\cdot\overline{g\left(t_{1}-\tau\right)}\right\} \left(\tau,\omega,t_{2}\right)\cdot g\left(t_{2}-\tau\right)\right\} \left(\tau,\omega\right)
}%
\end{equation}%

We can see that this formulation implies that the spectrogram is the
magnitude squared of the expected STFT spectrogram. First, remark
that%
\begin{equation}%
{
R\left(t_{1},t_{2}\right)=\mathbb{E}\left[X\left(t_{1}\right)\overline{X\left(t_{2}\right)}\right]=\int_{C}\int_{C}X\left(t_{1}\right)\overline{X\left(t_{2}\right)}\,dF_{X}\left(t_{1}\right)dF_{X}\left(t_{2}\right)
}%
\end{equation}%
where $F_{X}\left(t\right)$ is the cumulative distribution of $X$ at
time $t$ with support in $C$:%
\begin{align}%
S\left(\tau,\omega\right) & =\int_{-\infty}^{\infty}\int_{-\infty}^{\infty}\left(\int_{C}\int_{C}X\left(t_{1}\right)\overline{X\left(t_{2}\right)}\,dF_{X}\left(t_{1}\right)dF_{X}\left(t_{2}\right)\right)\cdot\nonumber \\
& \qquad\qquad\cdot\overline{g\left(t_{1}-\tau\right)}e^{-2j\pi t_{1}\omega}\,dt_{1}g\left(t_{2}-\tau\right)e^{2j\pi t_{2}\omega}dt_{2}\nonumber \\
& =\int_{C}\int_{C}\left[\int_{-\infty}^{\infty}X\left(t_{1}\right)\overline{g\left(t_{1}-\tau\right)}\,e^{-2j\pi t_{1}\omega}\,dt_{1}\right]\cdot\nonumber \\
& \qquad\qquad\cdot\left[\int_{-\infty}^{\infty}\overline{X\left(t_{2}\right)}g\left(t_{2}-\tau\right)e^{2j\pi t_{2}\omega}dt_{2}\right]\,dF_{X}\left(t_{1}\right)dF_{X}\left(t_{2}\right)\nonumber \\
& =\int_{C}\int_{C}\left[\int_{-\infty}^{\infty}X\left(t_{1}\right)\overline{g\left(t_{1}-\tau\right)}\,e^{-2j\pi t_{1}\omega}\,dt_{1}\right]\cdot\nonumber \\
& \qquad\qquad\cdot\overline{\left[\int_{-\infty}^{\infty}X\left(t_{2}\right)\overline{g\left(t_{2}-\tau\right)}e^{-2j\pi t_{2}\omega}dt_{2}\right]}\,dF_{X}\left(t_{1}\right)dF_{X}\left(t_{2}\right)\nonumber \\
& =\int_{C}\left|\int_{-\infty}^{\infty}X\left(t_{1}\right)\overline{g\left(t_{1}-\tau\right)}\,e^{-2j\pi t_{1}\omega}\,dt_{1}\right|^{2}\cdot\nonumber \\
& \qquad\qquad\cdot\,dF_{X}\left(t_{1}\right)\cdot\left(\int_{C}dF_{X}\left(t_{1}\right)\right)%
\end{align}%

By simplifying the expression, we obtain the relationship between the
spectrogram and the expectation of the STFT
spectrogram:%
\begin{align}%
S\left(\tau,\omega\right) & =\int_{C}\left|\int_{-\infty}^{\infty}X\left(t\right)\overline{g\left(t-\tau\right)}\,e^{-2j\pi t\omega}\,dt\right|^{2}\,dF_{X}\left(t\right)=\int_{C}\left|V_{g}\left(\tau,\omega\right)\right|^{2}\,dF_{X}\left(t\right)\nonumber \\
& =\mathbb{E}\left[\left|V_{g}\left(\tau,\omega\right)\right|^{2}\right]%
\end{align}%

\subsection{Wavelet spectrum}

A wavelet transform is defined as the linear transformation of a
function $x\left(t\right)$ into a series of rescaled and shifted
wavelet basis $\psi_{lk}$ \citep{StatisticalWavelets-Vidakovic-2009}
in relation to a index $l$ and a scale $k$:%
\begin{equation}%
{
W_{X}\left(l,k\right)=2^{-\nicefrac{l}{2}}\int_{-\infty}^{\infty}x\left(t\right)\psi_{lk}\left(2^{-l}t-k\right)dt
}%
\end{equation}%

In contrast with the Fourier transform, the wavelet transform decomposes
a time series $x\left(t\right)$ into a series of components in time
and frequency generated by manipulation of
$\psi_{jk}\left(\cdot\right)$.

Chiann et al.~defined the wavelet spectrum at a time index $\tau$ and
scale $k$, with respect to the wavelet mother $\psi$, as \citep[eq.
8]{StatisticalWavelets-Vidakovic-2009, WaveletTime-Chiann-1998}%
\begin{equation}%
{
\eta_{lk}=\sum_{u=-\infty}^{\infty}\gamma\left(\tau\right)\Psi_{lk}\left(u\right)
}%
\end{equation}%
where $\Psi_{lk}\left(u\right)$ is the wavelet
autocorrelation function:%
\begin{equation}%
{
\Psi_{lk}\left(u\right)=\sum_{t=0}^{\infty}\psi_{lk}\left(t\right)\psi_{lk}\left(t+\left|u\right|\right)
}%
\end{equation}%

In this paper, we will focus on the Morlet, or Gabor, wavelet kernel
given by%
\begin{equation}%
{
\psi_{lk}\left(t\right)=c_{0}\pi^{-\frac{1}{4}}e^{-\frac{1}{2}t^{2}}\left(e^{l\sigma t}-e^{-\frac{1}{2}\sigma^{2}}\right)
}%
\end{equation}%
where $\sigma$ is a parameter that controls the time and frequency
resolution of the transform, and the $c_{0}$ is the normalization
factor%
\begin{equation}%
{
c_{0}=\frac{1}{\sqrt{1+e^{-\sigma^{2}}-2e^{-\frac{3}{4}\sigma^{2}}}}
}%
\end{equation}%
Chiann proposed an unbiased estimator for $P_{lk}$
\citep{WaveletTime-Chiann-1998}, the wavelet
periodogram:%
\begin{equation}%
{
\hat{\eta}_{lk}=\left(\sum_{t=0}^{T-1}X_{t}\psi_{lk}\left(t\right)\right)^{2}
}%
\end{equation}%
This estimator is unbiased when
$\sum_{t=-\infty}^{\infty}\left(1+\left|t\right|\right)\gamma\left(t\right)<\infty$
\citep[Thm. 1]{WaveletTime-Chiann-1998}.

\section{Frequency-selective dyadic aggregated autoregressive (DASAR) model}

This section is dedicated to explain further details on our proposed
time-varying spectrum representation. In essence, the dyadic aggregated
autoregressive (DASAR) model of a real-valued signal
$x\left(t\right)$, with length $T$, is a time-frequency
representation where its time-varying spectrum
$S\left(t,\omega\right)$ (Equation~\ref{eq:tv-spectrum}) is
approximated as the weighted sum of $L$ levels with different
discontinuous spectra:%
\begin{equation}%
{
S_X\left(t,\omega\right)=\sum_{i=0}^{L}\rho\left(i,\omega\right)S^{\left(i\right)}\left(t,\omega\right)+\nu\left(t,\omega\right)
}\label{eq:DASAR}%
\end{equation}%
where $\nu\left(t,\omega\right)$ is an approximation error.

At a level $i$, $S^{\left(i\right)}\left(t,\omega\right)$ is a
piece-wise spectrum constructed using localized stationary spectra
estimated over a dyadic-division of the signal in the
time-domain:%
\begin{equation}%
{
S^{\left(i\right)}\left(t,\omega\right)=\begin{cases}
S_{0}^{\left(i\right)}\left(\omega\right)+\varepsilon_{0}\left(\omega\right) & 0\le t<\frac{T}{2^{i}}\\
S_{1}^{\left(i\right)}\left(\omega\right)+\varepsilon_{1}\left(\omega\right) & \frac{T}{2^{i}}\le t<2\frac{T}{2^{i}}\\
\vdots & \vdots\\
S_{2^{i}-1}^{\left(i\right)}\left(\omega\right)+\varepsilon_{2^{i}-1}\left(\omega\right) & T\frac{2^{i}-1}{2^{i}}\le t < T
\end{cases}
}%
\end{equation}%
where each piece-wise spectrum
$S_{j}^{\left(i\right)}\left(\omega\right)$ is a aggregated
second-order autogressive model $\text{ASAR}\left(k\right)$, i.e., the
weighted sum of $K$ stochastic oscillators described by a second-order
autogressive model $\text{SAR}\left(\omega,\tau\right)$ (with a
central frequency $\omega$, a randomness $\tau$ and a \emph{weight}
$\sigma_{k}^{\left(i,j,k\right)}$):%
\begin{equation}%
{
S_{j}^{\left(i\right)}\left(t,\omega\right)\sim\text{ASAR}\left(k\right)=\sum_{k=1}^{K}\text{SAR}\left(\omega^{\left(i,j,k\right)},\tau^{\left(i,j,k\right)},\sigma_{k}^{\left(i,j,k\right)}\right)
}\label{eq:ASAR-decomp}%
\end{equation}%

\begin{figure}
\includegraphics[width=1\textwidth,height=\textheight]{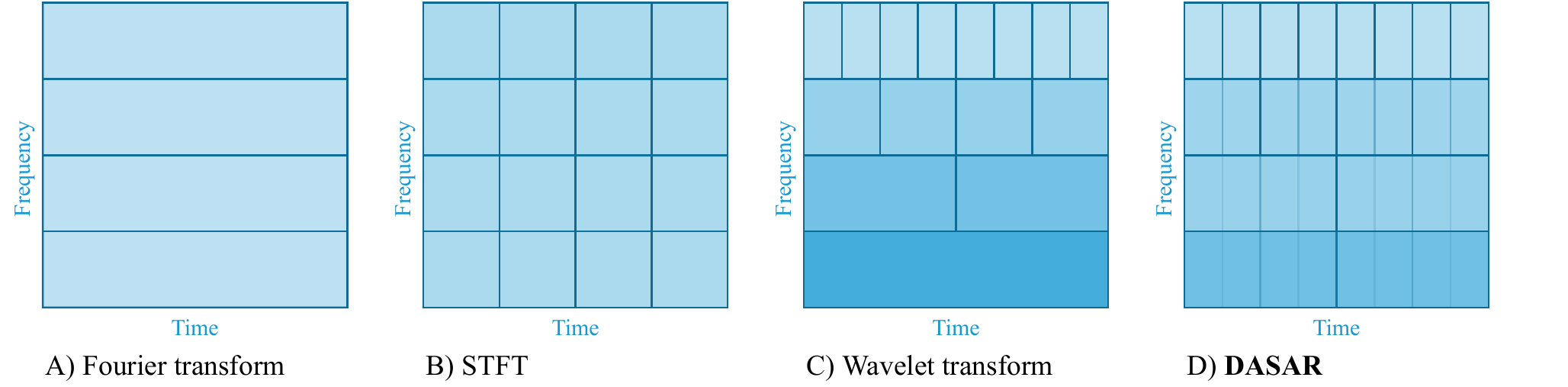}
\caption{Time-frequency division scheme in A) Fourier transform, B)
Short-time Fourier transform, C) Wavelet transform, D) Dyadic aggregated
autoregressive model (DASAR)}
\label{fig:comparison}
\end{figure}

A visual description of our time-frequency schematic division in
comparison with Fourier transform, STFT or wavelet transforms is shown
in Figure~\ref{fig:comparison}.

\subsection{Local stationarity approximation}

Mallat et al.~defined a property for approximating local stationary
processes by establishing local neighborhoods
\citep{AdaptiveCoVarEst-Mallat-1998}. Around a point of interest
$t^{*}$, let us define a neighborhood with a size
$l\left(t^{*}\right)$, i.e.~the time of analysis $t$ is limited to
the interval
$t\in\mathcal{T}=\left[t^{*}-\frac{l\left(t^{*}\right)}{2},t^{*}+\frac{l\left(t^{*}\right)}{2}\right]$.
Now, calculate the decorrelation length $d\left(\cdot\right)$ in the
neighborhood: the maximum distance $\Delta t$ in the region
$\mathcal{T}$ with a covariance is almost close to zero. Formally,
$d\left(t^{*}\right)=\max_{\Delta t}C\left(\tau,\Delta t\right),\,t,t^*\in\mathcal{T}$
such that $C\left(\tau,\Delta t\right)\le\varepsilon$ for a tolerance
error $\varepsilon$. When the decorrelation length is upper-bounded by
the neighborhood size:
$d\left(t^{*}\right)<\frac{1}{2}l\left(t^{*}\right)$, we can
approximate that the covariance $C\left(\cdot\right)$ by a stationary
covariance $C_{t^{*}}\left(\cdot\right)$:%
\begin{equation}%
{
C\left(\tau,\Delta t\right)\approx C_{t^{*}}\left(\Delta t\right)\quad\vert\Delta t\vert\le\frac{l\left(t^{*}\right)}{2}
}%
\end{equation}%

Therefore, in this local region, the spectrum is defined as a function
of the stationary covariance:%
\begin{align}%
S_{t^{*}}\left(\tau\right) & =\mathfrak{F}_{\nu\rightarrow\omega}\left\{ C_{t^{*}}\left(\nu\right)\right\} \left(\omega\right)=\int_{-\infty}^{\infty}C_{t^{*}}\left(\nu\right)e^{-j\omega\upsilon}d\upsilon%
\end{align}%

Hence, we will further approximate $S_{t^{*}}\left(\tau\right)$ using
an aggregate autoregressive model.

\hypertarget{subsec:SAR}{%
\subsection{Autoregressive models as stochastic
resonators}\label{subsec:SAR}}

Stochastic oscillations in biological signals can be effectively modeled
through autoregressive models
\citep{EffectElectrocardiogramInterference-Govindan-2016}. Let us start
modeling a single-resonator signal using an autoregressive model of second-order AR($2$), or SAR. 
This type of autoregressive model describes a
real and finite signal $x\left(t\right)$ (sampled at a fixed interval
$\nicefrac{1}{f_{s}}$) as the stochastic process generated by a linear
combination of the two previous points, $x\left(t-1\right)$ and
$x\left(t-2\right)$, added to an error term
$\varepsilon\left(t\right)$:%
\begin{equation}%
{
x\left(t\right)=\phi_{1}x\left(t-1\right)+\phi_{2}x\left(t-2\right)+\varepsilon\left(t\right)
}%
\end{equation}%
where $\phi_{1}$ and $\phi_{2}$ are known as the
autoregressive coefficients. In this case, without loss of generality,
we assume that $\mathbb{E}[X(t)]=0$ with the error component modeled
as a zero-mean random variable, with variance
$\sigma_{\varepsilon}^{2}$,
$\varepsilon\left(t\right)\sim\mathcal{WN}\left(0,\sigma_{\varepsilon}^{2}\right)$,
and serially uncorrelated
$\mathbb{E}\left[\varepsilon\left(t\right)\varepsilon\left(t-u\right)\right]=0\;\forall u$.

The spectrum $S_{x}\left(\omega\right)$ of this stationary SAR process
is characterized by:%
\begin{align}%
S_{x}\left(\omega\right) & =\frac{\sigma_{\varepsilon}^{2}}{\left|1-\phi_{1}e^{-2j\pi\omega}-\phi_{2}e^{-4j\pi\omega}\right|^{2}}
\label{eq:SAR}%
\end{align}%
where $\omega$ is the normalized frequency with respect to the
sampling frequency $f_{s}$: $\omega=\frac{f}{f_{s}}$.

We should note that $S_{x}\left(\omega\right)$ has a single maximum at
the dominating frequency $\omega^{*}$:

\begin{equation}%
{
\omega^{*}=\frac{1}{2\pi}\arccos\left(\frac{\phi_{1}\left(\phi_{2}-1\right)}{4\phi_{2}}\right)=\frac{1}{2\pi}\arctan\left(\frac{\phi_{2}}{\phi_{1}}\right)
}\label{eq:ar2-freq}%
\end{equation}%
when $\phi_{1}^{2}+4\phi_{2}<0$. Given our interest in the spectral
information contained in a SAR process, we can reformulate the
autoregressive coefficients based on $\omega^{*}$ and a parameter
$\tau$:%
\begin{equation}%
{
\phi_{1} =\frac{2}{1+e^{-\tau}}\cos\left(2\pi\omega^{*}\right)
}\label{eq:phi1}%
\end{equation}%
\begin{equation}%
{
\phi_{2} =-\frac{1}{\left(1+e^{-\tau}\right)^{2}}
}\label{eq:phi2}%
\end{equation}%
Therefore, SAR processes can be fully specified by the triplet
$\left(\omega,\tau,\sigma_{\varepsilon}\right)$. In these models,
$\tau$ controls the randomness of the central oscillation frequency:
small values can model signals with frequency components widely spread
around $\omega^{*}$, and large values of $\tau$ can model signals
with an clear central frequency.

Nevertheless, other autoregressive representations can have act as
randomness-controlled oscillation property as well. For instance, a
fourth-order autoregressive AR($4$) process given by
\begin{equation}%
{x\left(t\right)=\sum_{\ell=1}^{4}\phi_{4}x\left(t-\ell\right)+\varepsilon\left(t\right)}%
\end{equation}%
have only a single resonating frequency $\omega^{*}$ (that is also
described by Equation~\ref{eq:ar2-freq}) when the autoregressive
coefficients are defined as

\begin{equation}%
{
\left(
\begin{array}{c}
\phi_{1}\\
\phi_{2}\\
\phi_{3}\\
\phi_{4}
\end{array}\right)=\left(\begin{array}{c}
\frac{4}{\rho}\cos\left(\omega^{*}\right)\\
\frac{2}{\rho^{2}}\left(1-2\cos^{2}\left(\omega^{*}\right)-2\right)\\
\frac{4}{\rho^{2}}\cos\left(\omega^{*}\right)\\
-\frac{1}{\rho^{4}}
\end{array}
\right)
}%
\end{equation}%
where $\rho=1-e^{-\tau}$

AR($2$), or SAR, is the lowest-order autoregressive model that can
represent bandpass signals with a single maximum frequency. In order to
keep an efficient parameter estimation and a compact representation, in
this paper, we rely on the use of SAR models to represent stochastic
oscillation behaviors.

\hypertarget{sec:ASAR}{%
\subsection{Aggregated autoregressive model}\label{sec:ASAR}}

Biomedical signals, such as EEG
\citep{AtlasEEGPatterns-Stern-2013, ComputerizedEEGRvws-Lundgren-1990, EEGSigProcessing-Sanei-2007, EEGSigProcessing-Hu-2019, MethodsEEGSig-Al-Fahoum-2014, EvolutionaryStateSpaceModel-Gao-2016}
and fNIRS
\citep{InvestigateFreqSpectrum-NaderiKhojastehfar-2019, PhysFluctuationsShow-FernandezRojas-2017, ReliabilityFrontalEye-Yaramothu-2020}
are known to have several relevant frequency components that could be
associated with some physiological process. However, the above-mentioned
SAR models are limited to model single-frequency systems. Nevertheless,
it is possible to take advantage of the concise stochastic frequency
representation offered by SAR representations and extend it towards an
aggregated model, i.e., the superposition of several single-frequency
stochastic models.

A general aggregated model $\text{AAR}\left(p,K\right)$ is defined as
the sum of $K$ uncorrelated components where each one is characterized
through an $\text{AR}\left(p\right)$ process:%
\begin{equation}%
{
y\left(t\right)=\sum_{k=1}^{K}z_{k}\left(t\right)
}%
\end{equation}%

where each $z_{k}\left(t\right)$ is a latent unobserved
$\text{AR}\left(p\right)$ time series. Some theoretical properties of
these types of models were introduced by Chong et al.
\citep{TimeSeriesProp-Chong-2001} and generalized by Dacunha-Castelle et
al.~for $\text{AR}\left(p\right)$ processes
\citep{AggregationARProcesses-Dacunha-Castelle-2008}.

We restrict to SAR models for the representation of the latent processes
$z_{k}\left(t\right)$ in order to associate, each latent component
with a central frequency $\omega_{k}^{*}$, and a frequency randomness
$\tau_{k}$ (Equations~\ref{eq:phi1}, \ref{eq:phi2}, \ref{eq:SAR}). In
consequence, each latent process can be fully parametrized
by:%
\begin{equation}%
{
z_{k}\left(t\right)=\frac{\cos\left(2\pi\omega_{k}^{*}\right)}{1+e^{-\tau_{k}}}x\left(t\right)-\frac{1}{\left(1+e^{-\tau_{k}}\right)^{2}}z_{k}\left(t-1\right)+\varepsilon_{k}\left(t\right)
}%
\end{equation}%
where $\varepsilon_{k}$ is the additive white noise
$\varepsilon_{k}\sim\mathcal{WN}\left(0,\sigma_{\varepsilon k}^{2}\right)$.

By definition, $z_{i}\left(t\right)$ and $z_{j}\left(t\right)$ are
assumed to be uncorrelated for $i\ne j$. Therefore, the spectrum of an
$\text{AAR}\left(2,K\right)$ process is determined by%
\begin{align}%
S_{y}\left(\omega\right) & =\sum_{k=1}^{K}\frac{\sigma_{\varepsilon k}^{2}}{\left|1-\frac{\cos\left(2\pi\omega_{k}^{*}\right)}{1+e^{-\tau_{k}}}e^{-2j\pi\omega}+\frac{1}{\left(1+e^{-\tau_{k}}\right)^{2}}e^{-4j\pi\omega}\right|^{2}}
\end{align}%

Note that for a deterministic value $\alpha\in\mathbb{R}$, scaling the
latent component by $\alpha z_{k}\left(t\right)$ do not modify the
autoregressive coefficients, but only modify the variance of
$\varepsilon_{k}\left(t\right)$ by a factor $\alpha^{2}$. Therefore,
to ensure identifiability, we assume that a latent component
$z_k\left(t\right)$ has a unit variance, while
$\sigma_{\varepsilon k}^{2}$ is its associated weight
(Equation~\ref{eq:ASAR-decomp}).

For brevity of notation, in the following sections, we denoted
$\text{AAR}\left(2,K\right)$ processes as ASAR($K$) models
associated with a set of parameters
$\left\{ \phi_{1,1},\phi_{2,1},\ldots,\phi_{1,K},\right.$
$\left.\phi_{2,K},\sigma_{1}^{2},\ldots,\sigma_{K}^{2}\right\}$.

\hypertarget{sec:Estimation-alg}{%
\subsection{ASAR(K) estimation method}\label{sec:Estimation-alg}}

Estimation of the parameters in an aggregated autoregressive model is
still an open problem typically addressed with Kalman filters as the main estimation
algorithm. Wong et al.~developed an algorithm based on
expectation-maximization and Kalman filters to estimate four
second-order autoregressive models ($\text{ASAR}\left(4\right)$)
\citep{NonstationaryVar-Wong-2006}. Gao et al.~proposed using 
least-squares estimation enhanced with block resampling while relying on
Kalman filtering for iterative parameter estimation for an
$\text{ASAR}\left(4\right)$ model
\citep{EvolutionaryStateSpaceModel-Gao-2016}. In both cases, previous
knowledge of the signal to be analyzed was required in order to define a
fixed resonating frequency $\left\{ \omega_{k}^{*}\right\}$. In
\citep{DyadicAggregatedAR-Pinto-Orellana-2020}, Pinto et al.~introduced
an iterative heuristic approach to estimate a general
$\text{ASAR}\left(K\right)$ which is only restricted by the maximum number of
components or the desired approximation error.

In this paper, we introduce an improved estimation method for an
$\text{ASAR}\left(K\right)$ model that also estimates the latent
components of a signal iteratively that allows to track time-varying
frequency-localized changes. The following estimation algorithm relies
in three main conditions:

\begin{itemize}
\tightlist
\item
  The signal that has a fixed, and unknown, number of main resonating
  frequencies with a minimum separation $\Delta f$ among them, i.e,
  $|\omega_i - \omega_j|\le\Delta f \;\forall i\ne j$.
\item
  The spectrum of the signal $y(t)$ is considered appropriate with the
  approximation error is lower than $\epsilon^*$.
\item
  It is possible to explore only a particular region of the spectrum
  that has some biological interpretation, i.e.~the spectrum region of
  interest (SROI) $\mathcal{W}^{\left(0\right)}$. Focalized tracking
  in frequency can reduce the number of iterations in the estimation
  algorithm while provide relevant and interpretable information.
  However, under lack of apriori knowledge,
  $\mathcal{W}^{\left(0\right)}=\left[0,\pi\right]$.
\end{itemize}

Therefore, given $\Delta f$ and $\mathcal{W}^{\left(0\right)}$, we
propose the following estimation algorithm composed of three stages:

\begin{enumerate}
\def\labelenumi{\arabic{enumi}.}
\item
  \emph{Estimate the spectrum using a smoothed fast Fourier transform
  (FFT) of $y\left(t\right)$}. The spectrum is given by
  $S^{\left(0\right)}\left(\omega\right)=S_{X}^{FFT}\left(\omega\right)=\int_{-\infty}^{\infty}\mathbb{E}\left[y\left(t\right)y\left(t-\ell\right)\right]e^{-j\pi\omega}dt$
  and it can be estimated using a smooth version of the FFT. 
  We suggest the use of a median filter over the FFT of a zero-padded version of $y\left(t\right)$. Recall that zero-padding
  in time domain implies interpolating using a Dirichlet kernel in the
  frequency domain. Other smoothing filters can be used, but the
  obtained spectrum smoothness should not be excessive in order to avoid
  remove relevant frequency information.
\item
  \emph{Fit an AR(2) model that describes the dominating frequency
  $\omega^{*}$}:

  \begin{enumerate}
  \def\labelenumii{\arabic{enumii}.}
  \item
    \emph{Define the local neighborhood}. Let $\omega^{*}$ be the
    frequency where $S^{\left(k\right)}$ has a maximum value in the
    SROI $\mathcal{W}^{\left(k\right)}$:%
    \begin{equation}%
    {
    \omega^{*}=\min S^{\left(k\right)}\left(\omega\right),\quad\omega\in\mathcal{W}^{\left(k\right)}
    }%
    \end{equation}%
    Then, define a Gaussian window
    $B\left(\omega;\omega^{*}\right)$ around $\omega^*$ with a
    standard deviation equal to the frequency separation
    $\Delta f$:%
    \begin{equation}%
    {
    B\left(\omega;\omega^{*}\right)=\kappa_{B}\exp\left(-\left(\frac{\omega-\omega^{*}}{\Delta f}\right)^{2}\right)
    }%
    \end{equation}%
    where $\kappa_{B}$ is a normalization factor that
    ensures $\sum_{\omega\in W}B\left(\omega;\omega^{*}\right)=1$ in
    the discrete set of frequencies $W$ were estimated from the FFT
    algorithm. It is clear that $B\left(\omega;\omega^{*}\right)$
    denotes the local neighborhood of the dominating frequency.
  \item
    \emph{Find the optimum local SAR model}. Recall that the SAR
    spectrum Equation~\ref{eq:SAR} and $S^{\left(k\right)}$ are
    smooth, and therefore, with no discontinuities in their first
    derivative. Therefore, we can define an $L_{2}$-distance between
    them in the neighborhoud defined by $B$:%
    \begin{equation}%
    {
    L\left(\omega^{*},\tau\right)=\sum_{\omega}\left(S^{\left(k\right)}\left(\omega\right)B\left(\omega;\omega^{*}\right)-S\left(\omega;\omega^{*},\tau\right)\right)^{2}
      }%
      \end{equation}%
      The oscillation randomness $\hat{\tau}$ can be
    estimated by finding the minimum value of a loss function
    $L\left(\omega\right)$:%
    \begin{equation}%
    {
    \tau^{*}=\arg\min_{\tau}\left(L\left(\omega^{*},\tau\right)\right)
     }%
     \end{equation}%
  \end{enumerate}
\item
  \emph{Update the spectrum parameters}. The portions of spectrum that
  were not covered by the local SAR spectrum are updated for the
  following iteration:%
  \begin{equation}%
  {
  S^{\left(k+1\right)}\left(\omega\right)=\left\lceil S^{\left(k\right)}\left(\omega\right)-S\left(\omega;\omega^{*},\tau^{*}\right)\right\rceil ^{+}
  }%
  \end{equation}%
  and the SROI:%
  \begin{equation}%
  {
  \mathcal{W}^{\left(k+1\right)}=\mathcal{W}^{\left(k\right)}\backslash\left[\omega^{*}-\frac{1}{2}\Delta f,\omega^{*}+\frac{1}{2}\Delta f\right]
  }%
  \end{equation}%
\item
  \emph{Repeat until convergence} when either a) the error tolerance is
  reached
  $\sum_{\omega}\left|S^{\left(k+1\right)}\left(\omega\right)\right|<\epsilon^*$;
  or b) all the SROI was covered, i.e.,
  $\mathcal{W}^{\left(k+1\right)}=\emptyset$; or c) the maximum number
  of desired components have been estimated.
\end{enumerate}

\subsection{Dyadic time partition}

ASAR$\left(K\right)$ models provide a compact representation for
stationary signals with a finite number of main frequencies. However,
the model cannot be applied to non-stationary processes which frequency
properties are changing over time. Nevertheless, we can address this
issue by defining intervals where the local-stationarity property holds
(Equation~\ref{eq:tv-spectrum}) such as we approximate the time-varying
spectrum by piece-wise stationary spectra.

DASAR applies this concept by using a dyadic decomposition into levels
and segments in the time-domain. A similar approach was also used in
other non-stationary representations such as SLEX (smoothed localized
complex exponential basis) and derived methods
\citep{DiscriminationClassifNonstationary-Huang-2004, SLEXModelNonStationary-Ombao-2002, TimeFreqSpectralEst-Cranstoun-2002, ClusteringNonlinearNonstationary-Harvill-2017}.

For a level $i$, a signal $x\left(t\right)$ with $T$ points will
be split into $2^{i}$ segments: $j=0,1,..2^{i}-1$. Assuming local
stationarity in the segment $j$, the spectrum can be approximated
using an ASAR$\left(K\right)$ and estimated using the algorithm of
Section~\ref{sec:Estimation-alg}:%
\begin{equation}%
{
S_{j}^{\left(i\right)}\left(t,\omega\right)=\text{ASAR}\left(K\right)+\varepsilon_{i}\left(\omega\right)
}%
\end{equation}%
where $\varepsilon_{i}\left(\omega\right)$ is a
stationary zero-mean noise. Recall that this approximation is only
defined on the time interval
$t\in\left[k\frac{T}{2^{i}},\left(k+1\right)\frac{T}{2^{i}}\right)$.

Note that in consequence,
$S_{j}^{\left(i\right)}\left(t,\omega\right)$ is described by $K$
triplets of components
$\left\{ \left(\omega_{k},\tau_{k},\sigma_{k}^{2}\right)\right\} _{k=1}^{K}$
that associates a precise central frequency $\omega_{k}$, its
randomness $\tau_{k}$ and magnitude $\sigma_{k}^{2}$ of a $k$-th
component (Section~\ref{sec:ASAR}). We can improve the approximation of
$S_{j}^{\left(i\right)}\left(t,\omega\right)$ by choosing a higher
number of components $K$.

Thus, the spectrum (at a level $i$) in the interval
$t\in\left[0,T\right]$ is described by the piece-wise
function:%
\begin{equation}%
\protect\hypertarget{eq:DASAR-representation}{}{
S^{\left(i\right)}\left(t,\omega\right)=%
\begin{cases}
S_{0}^{\left(i\right)}\left(\omega\right)+\varepsilon_{0}\left(\omega\right) & 0\le t<\frac{T}{2^{i}}\\
S_{1}^{\left(i\right)}\left(\omega\right)+\varepsilon_{1}\left(\omega\right) & \frac{T}{2^{i}}\le t<2\frac{T}{2^{i}}\\
\vdots & \vdots\\
S_{2^{i}-1}^{\left(i\right)}\left(\omega\right)+\varepsilon_{2^{i}-1}\left(\omega\right) & T\frac{2^{i}-1}{2^{i}}\le t<T
\end{cases}
}\label{eq:DASAR-representation}%
\end{equation}%

The accuracy of this representation will be linked to the local
stationarity of the division. In wavelet transforms, the frequency
bandwidth of wavelet basis assume that larger segments are more suitable
to represent low-frequency spectra and shorter intervals more suitable
for modeling high-frequencies
\citep[p.~132]{WaveletTourSig-Mallat-2009}. We can apply a similar
principle and encode the information of each level using a non-negative
weight function $\rho\left(i,\omega\right)$ that quantifies the
assumed certainty of the DASAR spectrum approximation using $2^{i}$
segments at a frequency $\omega$.

The weight function $\rho\left(\cdot\right)$ should satisfy
$\sum_{i=1}^{L}\rho\left(i,\omega\right)=1\,\forall\omega\in\mathcal{W}^{\left(0\right)}$.
Then, we can construct the full DASAR$(L, K)$ spectrum as the sum of
the piece-wise spectra $S^{\left(i\right)}\left(t,\omega\right)$ for
each level $i$, but using the weight function
$\rho\left(\cdot\right)$ for penalizing their
contribution:%
\begin{equation}%
\protect\hypertarget{eq:DASAR-spectrum}{}{
S_X\left(t,\omega\right)=\sum_{i=0}^{L}\rho\left(i,\omega\right)S^{\left(i\right)}\left(t,\omega\right)+\nu\left(t,\omega\right)
}\label{eq:DASAR-spectrum}%
\end{equation}%
where $\nu\left(t,\omega\right)$ is the inherent approximation error.

\section{Data validation}

DASAR provides a robust framework for capturing time-varying spectral properties in a time series. In this section, we validate the model with synthetic and realistic biomedical data. We compared DASAR spectrum estimations to short-time Fourier transform (STFT) spectrograms and the wavelet transform (WT) scalograms, which served as a baseline for comparison. The synthetic data enable us to compare DASAR's ability to identify spectral changes defined \textit{a priori}. Additionally, analysis with real biomedical signals would allow us to evaluate the method's potential to generate novel descriptive features.

\subsection{Synthetic experiment with chirp signals}

The first validation scenario for DASAR is a non-stationary model where the ground truth is known. For our purposes, we choose to use a chirp signal as a synthetic model in which transient patterns denote extremely non-stationary frequency modulation effects. This type of signals has been recognized as a plausible model for some physical phenomena \citep[p.10-11]{ExplorationsTimeFreq-Flandrin-2018}, as well as physiological processes where it has been identified as a subject-dependent epileptic EEG biomarker \citep{BrainChirpsSpectrographic-Schiff-2000}.

In our comparison analysis, we simulate a signal consisting of a 4 Hz sinusoidal component and a sinusoidal resonator with a frequency that linearly increases from 1 Hz to 3 Hz over time:
\begin{equation}%
{ y\left(t\right)=5\sin\left(8\pi t\right)+5\sin\left(8\pi t\left(1+0.01t\right)\right)+\frac{1}{2}\varepsilon\left(t\right) }%
\end{equation}%
where $\varepsilon\left(t\right)$ is a zero-mean white noise with a unit-variance.

\subsection{Real experiment: concurrent EEG-fNIRS signals}

The second validation phase involves evaluating DASAR's performance on real biomedical signals. In order to accomplish this, we gathered signals using a custom-built instrument capable of recording high-quality concurrent EEG-fNIRS.

\begin{figure}
\includegraphics[width=1\textwidth,height=\textheight]{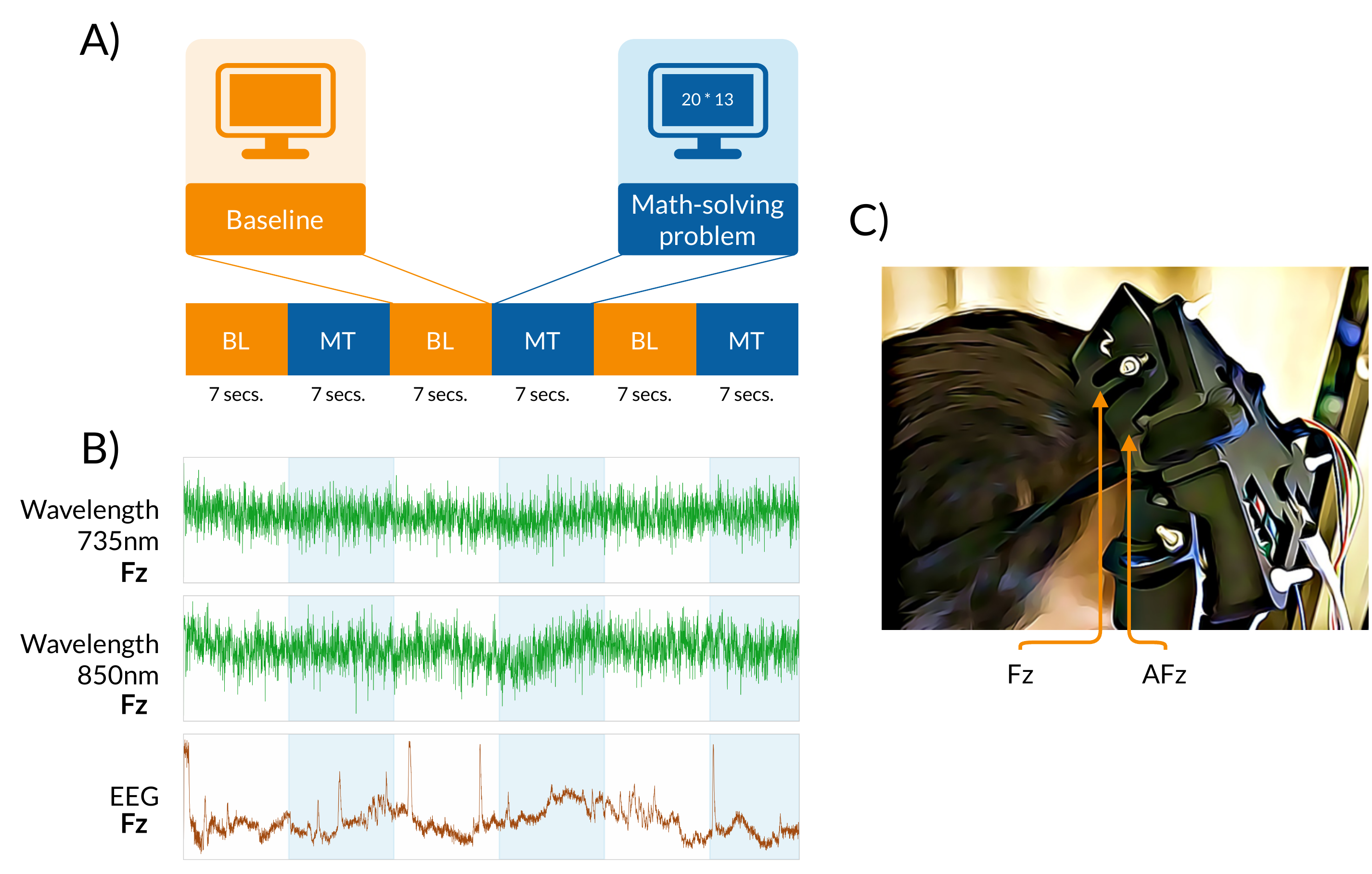}
\caption{Experiment setup for real signal acquisition: A) experiment
protocol; B) recorded biomedical signals (in normalized units); C)
hybrid EEG-fNIRS montage layout.}
\label{fig:montage}
\end{figure}
\def\registered{{\textregistered}}

The custom-built device consists of three modules: hemodynamic, biopotential, and control that capture and preprocess signals. The biopotential module's reliance on the TI-ADS1299 integrated circuit \citep{ADS1299LowNoise8Channel-TexasInstruments-2017, FundamentalsPrecisionADC-Lizon-2020} ensured EEG signal quality and reliability. Rashid et al. stated that, when combined with supplementary circuitry, this component may provide signal quality comparable to clinical-level EEG systems such as NuAmps (Compumedics Neuroscancopyright, Dresden, Germany) \citep{EEGExptudy-Rashid-2018}. Furthermore, the hemodynamic module produces fNIRS signals, with a high signal-to-noise ratio, at three different wavelengths: 735nm, 805nm, and 850nm. The control module implemented in an integrated digital signal processing board (DSP) performs the complementary functions that interconnect and manage both modules. We refer to \citep{SHADEAbsorption-Sherkat-2020} for further details relating to the instrumentation. 

This device was mounted, ensuring that the light source is located on the channel AFz, according to the extended 10-20 positioning system, while three photodiodes positioned at FP1, FP2, and Fz (Figure~\ref{fig:montage}.C). This arrangement of light sources and detectors provides nine fNIRS channels that map the hemodynamic response on the frontal and prefrontal lobe. 

EEG-fNIRS signals were collected from a single subject in the prefrontal cortex during a mental arithmetic task experiment. This family of neuropsychological activities has been extensively studied employing EEG and fNIRS
\citep{BrainPotentialsMental-Pauli-1994, RelationAsymmetryPrefrontal-Tanida-2004, RTMentalArithmetic-Wang-2013, CorticalOxygenConsumption-Verner-2013}.
The experiment used in this paper includes six sections alternating
between arithmetic tasks and baseline intervals, where each section had an
identical span of seven seconds (Figure~\ref{fig:montage}). During the
task period, the subject was asked to perform a two-digit subtraction,
while during the baseline interval, only a cross sign was shown on the
screen to keep the subject's attention. Concentration changes in HbO and
HbR were estimated from the measured light intensity of the three wavelengths
using the modified Beer-Lambert law (MBBL). Two parameters
of the MBBL: differential path length and extinction coefficients, were
estimated using the Scholkmann model
\citep{GeneralEquationDifferential-Scholkmann-2013}, and the
Gratzer-Kollias model \citep{TabulatedMolarExtinction-Prahl-1998},
respectively. Note that all the signals were recorded at a 100 Hz
sampling rate.

\section{Results and discussion}

\begin{figure}
\includegraphics[width=1\textwidth,height=\textheight]{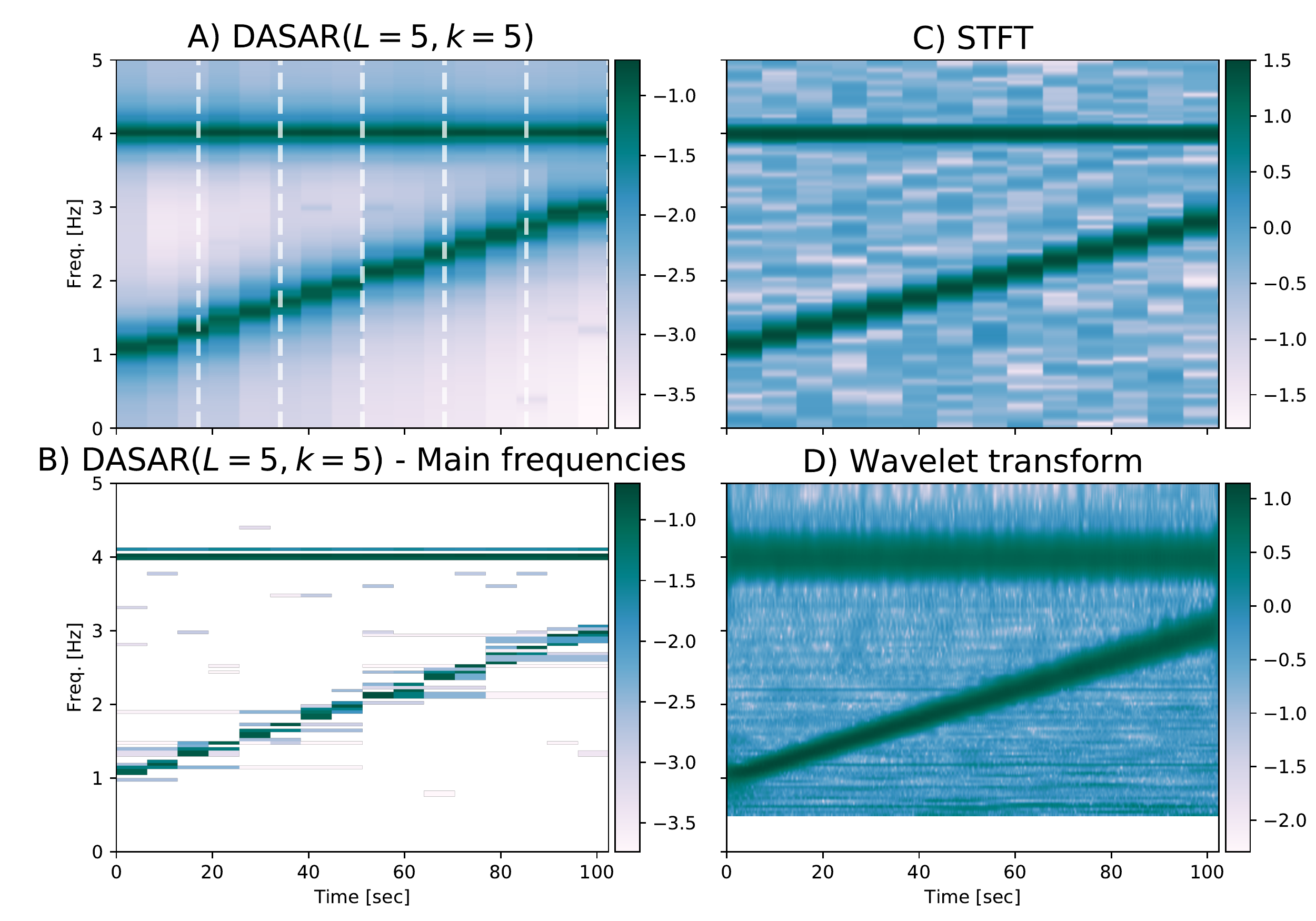}
\caption{Comparison among time-frequency representations of a
nonstationary signal: A) DASAR$(L=5,K=5)$ spectrum, B) Central frequency distribution of the
DASAR$(L=5,K=5)$ model, C) Short-time Fourier transform (STFT), D) Wavelet
transform. Note that magnitudes are expressed in decibels.}
\label{fig:results-Chirp}
\end{figure}

\begin{figure}
\includegraphics[width=1\textwidth,height=\textheight]{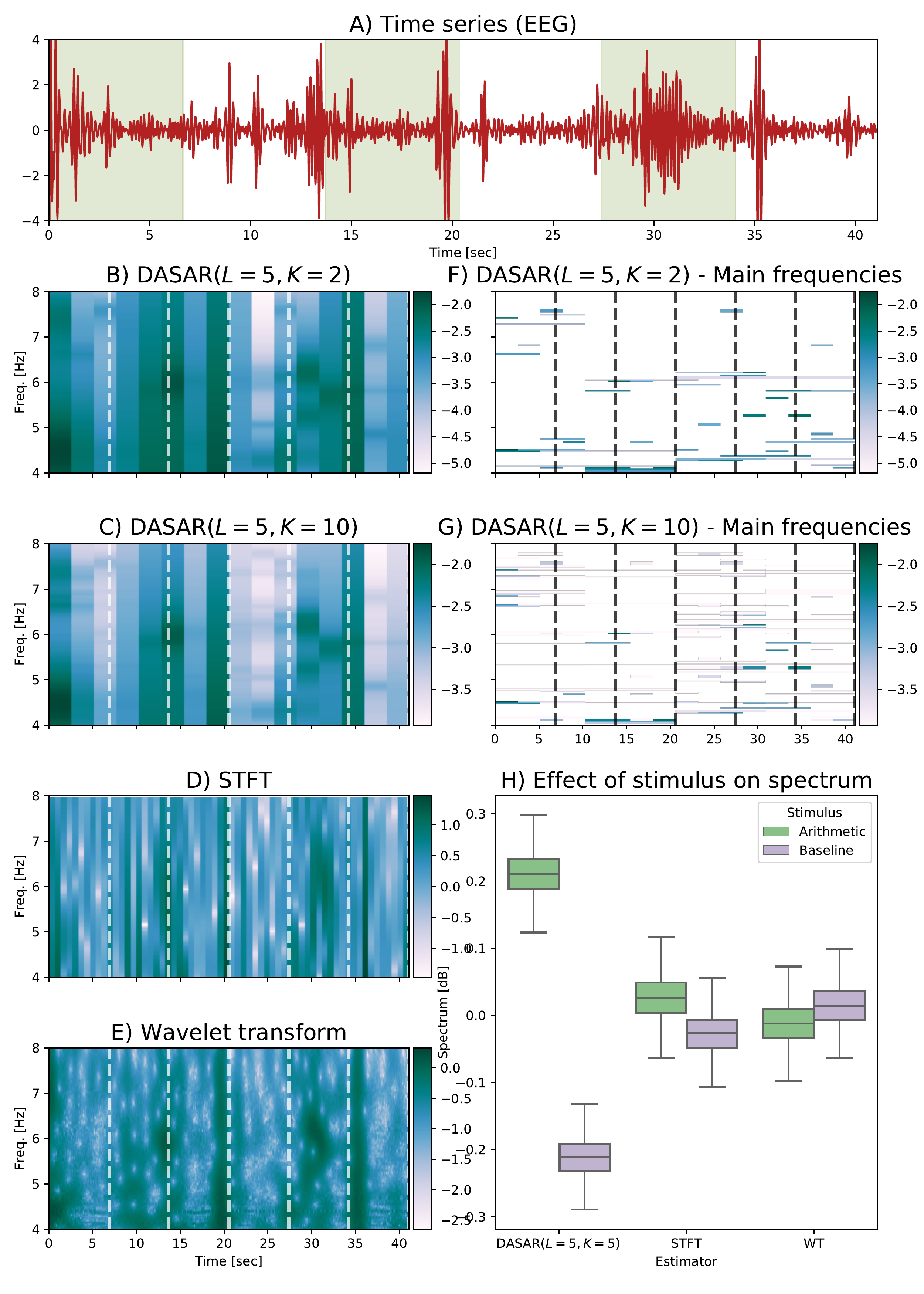}
\caption{EEG time-frequency representations in the theta band: A)
DASAR$(L=5,K=2)$ and B) DASAR$(L=5,K=10)$, C) STFT, D) WT, E) main
frequencies in DASAR$(L=5,K=2)$, and F) DASAR$(L=5,K=10)$, and G)
comparison of the bootstrap spectrum means between DASAR$(L=5,K=5)$,
STFT and WT.}
\label{fig:results-EEG}
\end{figure}

\begin{figure}
\includegraphics[width=1\textwidth,height=\textheight]{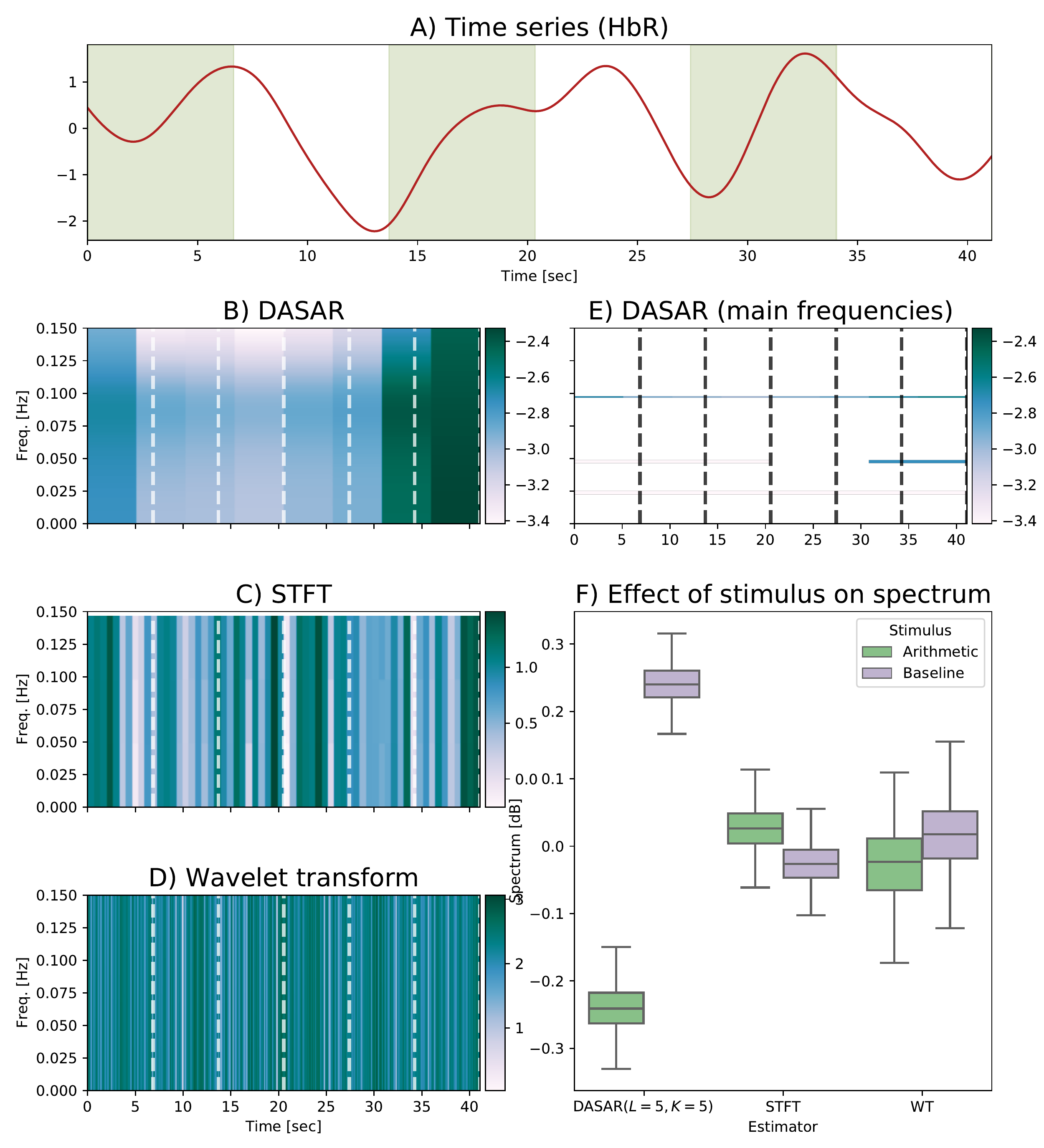}
\caption{fNIRS-HbR time-frequency representations in the interval
0-100 mHz: A) DASAR$(L=5,K=5)$, B) STFT, C) WT, D) main frequencies in
DASAR$(L=5,K=5)$, and F) their comparison of the bootstrap spectrum
means.}
\label{fig:results-HbR}
\end{figure}

\begin{figure}
\includegraphics[width=1\textwidth,height=\textheight]{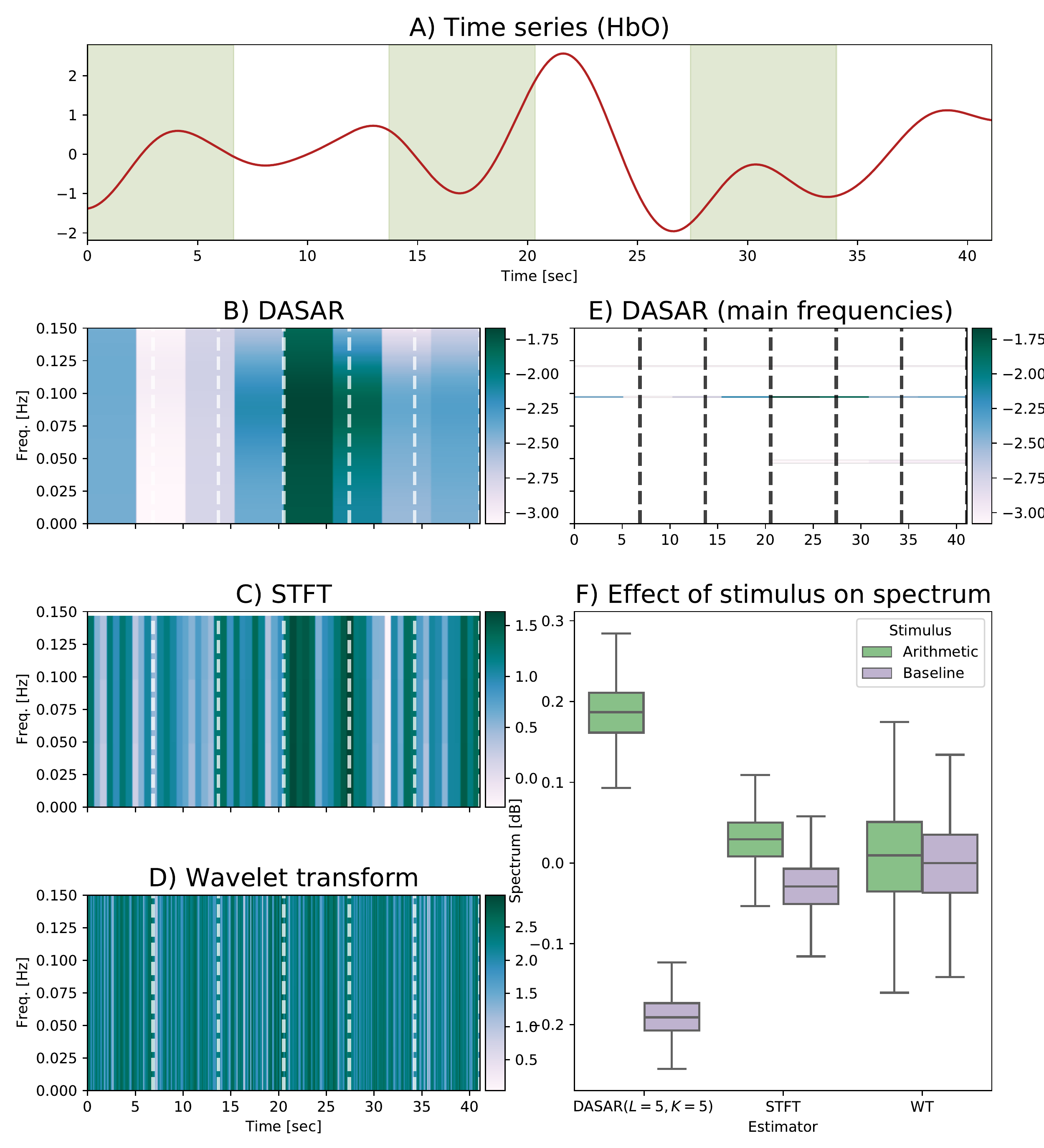}
\caption{fNIRS-HbO time-frequency representations in the interval
0-100 mHz: A) DASAR$(L=5,K=5)$, B) STFT, C) WT, D) main frequencies in
DASAR$(L=5,K=5)$, and F) their comparison of the bootstrap spectrum
means.}
\label{fig:results-HbO}
\end{figure}

\subsection{Synthetic chirp signal}

A comparison between the estimates of the Wavelet scalogram, the STFT
spectrogram, and the DASAR spectrogram applied on the synthetic data is
shown in Figure~\ref{fig:results-Chirp}. The three methods, as it can be
observed, capture the time-varying frequency features of the chirp
signal properly (Figure~\ref{fig:results-Chirp}.A-C). As expected, the
level of precision of that identification varies according to the
applied method. In comparison with STFT, DASAR spectrogram performs
automatic filtering of the spurious components with a similar level
achieved by the Wavelet scalogram.

A relevant feature that was outlined in this experiment is
the intrinsic ability of DASAR to identify and track frequency-changing
oscillators. Recall that in the DASAR interpretation of the time-varying
spectrum, the spectrum at a time $t$ is the consequence of a finite
sum of oscillating components (that could suffer some randomness in
their frequency). Therefore, it is possible to visualize only those
critical components to evaluate their spectrum evolution in the signal
(Figure~\ref{fig:results-Chirp}.D). In contrast with STFT and Wavelet,
DASAR allows us to identify those main components and track them without
further processing: DASAR$(5, 5)$ was able to detect the oscillator
properly with a fixed frequency of 4 Hz and the component with varying
frequency from 1-3 Hz (Figure~\ref{fig:results-Chirp}.B). The only
required assumption, in this example, was a maximum of five components
per level, i.e., a combination of 25 possible frequencies per dyadic
interval.

\subsection{DASAR-EEG spectrogram}

In the analysis of electroencephalographic signals, we focus on the EEG
theta band, i.e., the 4-8 Hz frequency range. This brainwave subset 
seems to be correlated with cognitive arithmetic-related tasks as it was
found in problem-solving activities \citep{EEGThetaWaves-Schacter-1977},
Sternberg memory scanning \citep{FSignificanceTheta-Klimesch-2005},
two-digit addition \citep{RegionalHemo-Sammer-2007}, and two-digit
multiplication problems \citep{OscillatoryEEGCor-Grabner-2012}. The
comparison result with STFT, wavelet, and DASAR is summarized in
Figure~\ref{fig:results-EEG}.

We must recall that the DASAR model provides time-frequency
representations with different accuracy levels determined by its
parameters (number of levels $L$ and oscillatory components per
segment $K$). In this dataset, two possible time-frequency
representations were used: DASAR$(L=5,\,K=2)$
(Figure~\ref{fig:results-EEG}.B,F) and DASAR$(L=5,\,K=10)$
(Figure~\ref{fig:results-EEG}.H,G). DASAR$(L=5,\,K=2)$ denoted only the
two most relevant components capable of modeling each of the $2^L=32$
segments in which the time series is split.

Note that DASAR models were able to analyze the spectrum dynamics that
emerged exclusively within the spectrum region of interest SROI
$\mathcal{W}^{(0)}$. While similar constraints can be introduced to
wavelet transforms using an appropriate set of wavelet scales, this
mechanism is not straightforward. In the current experiment, both DASAR
models were constrained to $\mathcal{W}^{(0)}=[4,8]$ Hz.

DASAR$(L=5, K=2)$ revealed a different estimate of the time-varying
EEG spectrum (opposed to STFT and WT) with no exact main component but a
high-density region in the lower theta-band: 4-6 Hz. In addition, the
DASAR$(L=5, K=10)$ model (with an increased number of maximum
components) was used to assist in identifying the most dominant
oscillation. It showed that a faint oscillator at \textasciitilde4.102 Hz
was present during the whole session. No similar information could be
drawn from the alternative time-frequency models.

Finally, we evaluated the consistency of the conclusions inferred from
the three time-frequency models (DASAR, STFT, and WT) with the findings
reported in the biomedical literature. For instance, Sammer et al.~and Grabner et
al. described an increase of the theta-band power spectrum during
arithmetic tasks
\citep{RegionalHemo-Sammer-2007, OscillatoryEEGCor-Grabner-2012}. To
provide a similar metric, we standardized the magnitudes of each
spectrogram (or scalogram) during the whole session, and we compared the 
mean spectrum for each period (arithmetic and baseline) 
using the boostrap method (Figure~\ref{fig:results-EEG}.H). All time-frequency
methods demonstrated an increment in the average power spectrum when the
participant performed an arithmetic task. However, DASAR$(L=5, K=5)$
has exhibited the most prominent spectrum difference between both conditions.

\subsection{DASAR-fNIRS spectrogram}

In this paper, fNIRS data analysis was restricted to those
frequency components within the range 0-150 mHz
\citep{NoninvasiveInfraredMonitoring-Jobsis-1977, CerebralBlood-Hori-2014}.
This region contains biomarkers associated with autoregulatory functions
and, therefore, has a biological interpretation
\citep{CerebralBlood-Hori-2014}. Similar to the previous analysis, we
contrasted DASAR, STFT, and WT, and the results are compiled in
Figure~\ref{fig:results-HbO} (HbO) and Figure~\ref{fig:results-HbR}
(HbR).

Frequency variations on very-low-frequency intervals are very
challenging to be detected using STFT or wavelet scalograms
(Figure~\ref{fig:results-HbR}.C-D and Figure~\ref{fig:results-HbO}.C-D).
However, some patterns were discerned using a DASAR($L=5, K=5$) model:
two principal components have been identified throughout most of the
experiment. The strongest widespread oscillator had a central frequency
in the range \textasciitilde97.4-97.5 mHz,and the second oscillator
was located at \textasciitilde48.7 mHz. 
Note that in the latter, the HbR spectrogram power is higher than in HbO.
On the other hand, the first oscillator is strongly correlated to Mayer
waves, i.e., spontaneous biological artifacts with a similar frequency
originated due to arterial blood pressure \citep{MayerWavesReduce-Yucel-2016}.
The remarkable aspect of recognizing the exact subject-dependent oscillatory frequency of the 
Mayer waves is the possibility of removing its effect in the signal without 
disrupting other physiological components at close frequencies using an anti-notch filter. 

Furthermore, DASAR can provide a time-frequency representation with a
small set of parameters similar to wavelet packets. In our fNIRS
analysis, it is noteworthy that the time-varying parameters of two key
stochastic oscillators (at the frequencies mentioned above) may be
enough to explain the main information denoted in the spectrum.

Despite this information compactness, DASAR cannot ensure
smooth spectral transitions across time. However, once the model is fitted,
a continuous smooth spectral representation
$\hat S_X\left(t,\omega\right)$ (Equation~\ref{eq:DASAR}) for any
normalized frequency $\omega\in[0,\nicefrac{1}{2}]$ is defined at
any time point. As a consequence, DASAR does not need any additional
interpolation technique as would be required in other methods that
used~an FFT-based estimation method, such as SLEX
\citep{SLEXModelNonStationary-Ombao-2002} or STFT.

\section{Conclusion}

We introduced the dyadic aggregated autoregressive (DASAR) model as a
robust time-frequency analysis technique. This method captures the
time-fluctuating spectrum characteristics of a signal by independently
modeling the spectrum of signal segments with an aggregated
autoregressive model. Each autoregressive component in those aggregated
segment-level models can model a single stochastic oscillator, i.e.,
signal sources whose frequency is assumed to be unstable but located
around a finite mean value. As it was shown, the lowest-degree
autoregressive model that satisfies this single-oscillator property is
the second-order that DASAR uses by design (it should be noted
that the proposed model is not restricted to this model order). Additionally
to this structure, we have introduced a spectrum region of interest
(SROI) where the time-varying spectrum estimation process will be focused. 
As a result, SROI can provide accurate estimates in the
frequency intervals where biomedical signals are recognized to have a particular
interpretation.

DASAR was applied to two types of biomedical signals:
electroencephalogram (EEG) and functional near-infrared spectroscopy
(fNIRS) time series during a two-digit multiplication task. Both types of
signals were analyzed in a separate spectral domain: the EEG processing 
was focused on the theta-band: 4-8 Hz, while the fNIRS analysis was 
concentrated in a very-low-frequency range (10-100 mHz). 
In both types of signals, it was observed that DASAR 
increased the contrast of the mean time-varying 
spectrum, between both stimuli, in comparison with the
short-time Fourier transform (STFT) and complex Morlet wavelet (WT). 
DASAR also was able to detect subtle time-varying characteristics, such 
as the Mayer waves that were often neglected by the alternative methods 
due to its frequency scale.

From the observed results, we can emphasize three main unique
characteristics of DASAR in relation to STFT and WT. First, while DASAR
can estimate a similar time-frequency response as the above-mentioned methods,
the spectrum mean values in the fitted $ASAR(K)$ models can be used to precisely identify and
track the main oscillating components. Second, DASAR allows
interpretable frequency-focused tracking in the spectrum region of
interest. Finally, DASAR provides coverage of the entire time-varying 
spectrum representation with a reduced number of parameters. 
Based on this characteristics on the simulated data and the realistic
biomedical signals, we can suggest DASAR as an alternative time-frequency 
method to analyze hemodynamic and electrical time series.

\bibliographystyle{elsarticle-num}
\bibliography{Library,ExtraBib}

\end{document}